\documentclass[conference]{IEEEtran}
\usepackage[utf8]{inputenc}
\usepackage[english]{babel}
\usepackage{algorithm}
\usepackage{algorithmicx}
\usepackage{algpseudocode}
\usepackage{amsmath}
\usepackage{amssymb}
\usepackage{array}
\usepackage{arydshln}
\usepackage{bm}
\usepackage{booktabs}
\usepackage{braket}
\usepackage{cite}
\usepackage{hyperref}
\usepackage{enumitem}
\usepackage{graphicx}
\usepackage[mathscr]{euscript}
\usepackage{mathtools}
\usepackage{multirow}
\usepackage{nicefrac}
\usepackage{pgfplots}
\usepackage{qtree}
\usepackage{adjustbox}
\usepackage{rotating}
\pgfplotsset{compat=1.14}
\usepackage{stfloats}
\usepackage{subcaption}
\usepackage{url}
\usepackage{tikz}
\usetikzlibrary{shapes, backgrounds,calc, snakes}
\usetikzlibrary{decorations.pathreplacing}

\DeclarePairedDelimiter\smallfloor{\big\lfloor}{\big\rfloor}
\DeclarePairedDelimiter\floor{\big\lfloor}{\big\rfloor}
\tikzset{
	solid node/.style={circle,draw,inner sep=1.2,fill=black},
	hollow node/.style={circle,draw,inner sep=1.2}
}

\makeatletter
\def\BState{\State\hskip-\ALG@thistlm}
\algdef{SE}{Begin}{End}{\textbf{begin}}{\textbf{end}}
\makeatother


\algdef{SE}[DOWHILE]{Do}{DoWhile}{\algorithmicdo}[1]{\algorithmicwhile\ #1}%

\newcommand{\sref}[1]{Section~\ref{#1}}

\begin{document}

\title{\LARGE{SWAN: Swarm-Based Low-Complexity Scheme for PAPR Reduction}}

\author{\IEEEauthorblockN{Luis F.~Abanto-Leon\IEEEauthorrefmark{2},
					 	  Gek Hong (Allyson) Sim\IEEEauthorrefmark{2}, 
					 	  Matthias Hollick\IEEEauthorrefmark{2}, 
						  Amnart Boonkajay\IEEEauthorrefmark{3} and
						  Fumiyuki Adachi\IEEEauthorrefmark{4}
						}
\IEEEauthorblockA{\IEEEauthorrefmark{2}Secure Mobile Networking (SEEMOO) Lab, Technical University of Darmstadt, Department of Computer Science, Germany \\
\IEEEauthorrefmark{3}Institute for Infocomm Research (I2R), Singapore \\
\IEEEauthorrefmark{4}Wireless Signal Processing Research Group, ROEC, Tohoku University, Japan \\
Email:
\IEEEauthorrefmark{2}\{labanto,asim,mhollick\}@seemoo.tu-darmstadt.de,
\IEEEauthorrefmark{3}amnart{\_}boonkajay@i2r.a-star.edu.sg,
\IEEEauthorrefmark{4}adachi@ecei.tohoku.ac.jp}
}

\markboth{IEEE Transactions on Vehicular Technology,~Vol.~XX, No.~X, September~2016}{Shell \MakeLowercase{\textit{et al.}}:xxxxxxx}

\maketitle

\begin{abstract}
Cyclically shifted partial transmit sequences (CS-PTS) has conventionally been used in SISO systems for PAPR reduction of OFDM signals. Compared to other techniques, CS-PTS attains superior performance. Nevertheless, due to the exhaustive search requirement, it demands excessive computational complexity. In this paper, we adapt CS-PTS to operate in a MIMO framework, where singular value decomposition (SVD) precoding is employed. We also propose \texttt{SWAN}, a novel optimization method based on swarm intelligence to circumvent the exhaustive search. \texttt{SWAN} not only provides a significant reduction in computational complexity, but it also attains a fair balance between optimality and complexity. Through simulations, we show that \texttt{SWAN} achieves near-optimal performance at a much lower complexity than other competing approaches.      
\end{abstract}


\IEEEpeerreviewmaketitle

\section{Introduction}

The adoption of orthogonal frequency division multiplexing (OFDM) by various communication standards (e.g., WiFi, ISDB-T, LTE, 3GPP Rel. 15/16) stems from its capability to provide high data rates, augmented spectral efficiency, and robustness to multi-path fading \cite{b1}. However, OFDM signals suffer from a high peak-to-average power ratio (PAPR) \cite{b3} caused by the constructive combination of modulated subcarriers. OFDM signals with high PAPR are power-inefficient \cite{b4} and prone to distortion due to the non-linearity of radio frequency (RF) amplifiers. Distortionless amplification can be achieved by reducing the signal power (i.e., back-off mechanism), thus forcing the amplifier to operate in the linear amplification region. However, this procedure compromises the RF amplifier energy efficiency. Therefore, it is essential to develop new approaches without resorting to back-off mechanisms.


\textit{Literature review:} To reduce the PAPR, several approaches have been proposed. Clipping \cite{b4,b22, b23, gacanin2006:reduction-amplitude-clipping-level-ofdm-tdm} limits the signal amplitude to a maximum threshold, thus preventing large peaks but causing distortion and bit error rate (BER) degradation \cite{b5}. Companding \cite{b6, b27, b28} consists of compression at the transmitter (to avoid distortion) and signal expanding at the receiver (to restore the amplitude). However, the latter process also magnifies small-valued noise, thus causing BER degradation. Besides, tone reservation (TR) \cite{b26, b7} and tone injection (TI) \cite{b7, b24, b25} are techniques that can reduce the PAPR without affecting the BER performance. TR uses a subset of subcarriers for exclusively canceling large signal peaks. TI expands the conventional PSK/QAM constellations such that each symbol can be mapped into one of several possible representations, and the best symbol mapping that minimizes the PAPR is chosen for transmission.

Another subgroup of techniques suppresses the large peaks by applying phase rotations (at the transmitter) and phase de-rotation (at the receiver), which has the advantage of preserving the BER performance. The most representative techniques of this kind are selected mapping (SLM) \cite{b29, b8} and partial transmit sequences (PTS) \cite{b9, b31}. In SLM, each modulated subcarrier is altered by a phase rotation whereas, in PTS, the modulated subcarriers are divided into disjoint partitions, and each partition is affected by the same phase rotation. While SLM relies on the design of codebooks, PTS focuses on finding the optimal phase rotations from a set of admissible values. The performance of both approaches are similar, and their computational complexities are high. Cyclically shifted partial transmit sequences (CS-PTS) \cite{b10, b32, b33, b34} has not received much attention despite being superior to PTS and SLM. CS-PTS leverages the idea of PTS but incorporates additional time-domain cyclic shifting, which provides another degree of freedom that enables per-subcarrier phase rotation. This improves PAPR reduction but causes substantial complexity increase as more parameters have to be optimized (e.g., phase rotations and time shifts). 


\textit{Contributions:} To address the high search complexity of CS-PTS, we propose \texttt{SWAN}, a swarm-based optimization approach. \texttt{SWAN} controls the number of evaluations of potential solutions, thus maintaining the search complexity affordable with negligible impact on the optimality. Swarm-based approaches are characterized by exploration and exploitation. \emph{Exploration} is the capability of effectively sampling the search space without inspecting every possibility exhaustively. \emph{Exploitation} is the ability to capitalize on information obtained in previous iterations to produce more suitable solutions. Cuckoo search algorithm (\texttt{CSA}) \cite{b15} is a swarm-based approach inspired in the parasitic breeding behavior of some birds. It was shown though extensive experimentation that \texttt{CSA} outperforms other methods such as genetic algorithms (\texttt{GA}) \cite{b13} and particle swarm optimization (\texttt{PSO}) \cite{b12}. \texttt{CSA} has remarkable exploration capability, which is attributed to the usage of Levy flights. Nevertheless, the exploitation property of \texttt{CSA} is limited. We found that by improving the exploitation capability, the convergence rate of \texttt{CSA} could be substantially accelerated. Our proposed approach \texttt{SWAN} is an improvement to \texttt{CSA}, wherein we incorporate four additional mechanisms to \emph{(i)} boost the exploitation capability of \texttt{CSA} and \emph{(ii)} achieve a fair balance between exploration and exploitation. The following summarizes our contributions:

\begin{itemize}[noitemsep, topsep = 0pt, leftmargin = 0.4cm]
	\item CS-PTS has only been used in SISO systems \cite{b10, b32, b33, b34}. We are the first to adapt CS-PTS to operate in MIMO systems with singular valued decomposition (SVD) precoding. 
	\item We generalize the application of PTS and SLM to SVD-MIMO systems.
	\item We propose a novel swarm-based approach, \texttt{SWAN}, which finds near-optimal solutions (i.e., low-PAPR signals) at enhanced convergence rate and affordable complexity.
\end{itemize}

\section{Generalizing CS-PTS from SISO to SVD-MIMO}
\begin{figure*}[t!]
	\centering
	\includegraphics[width = 0.82\textwidth]{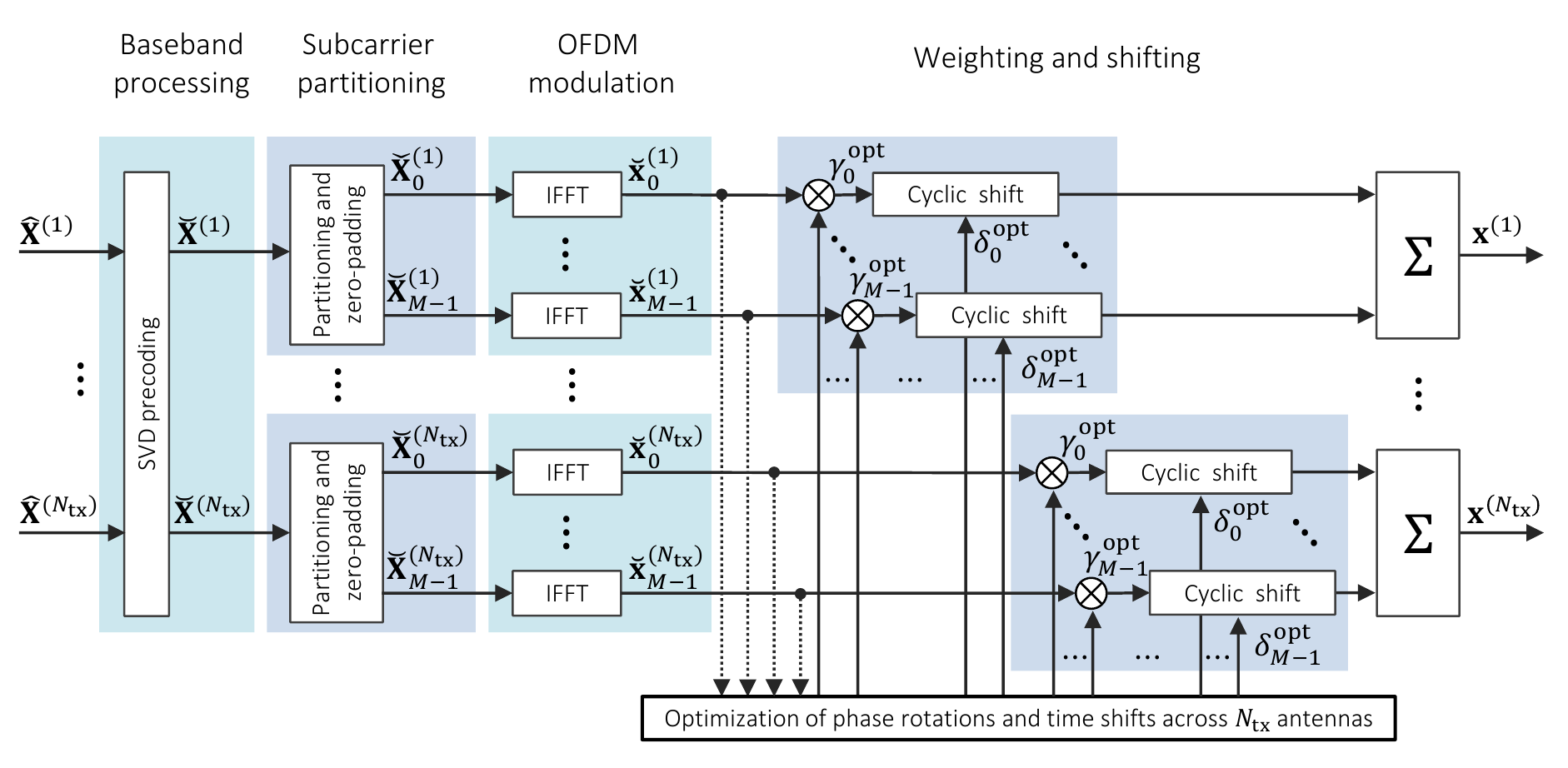}
	\caption{CS-PTS for a SVD-MIMO system}
	\vspace{-2mm}
	\label{f1}
	\vspace{-2mm}
\end{figure*}

Fig. \ref{f1} shows the implementation of CS-PTS for a SVD-MIMO system with $ N_c $ subcarriers and $ N_\mathrm{tx} $ antennas at the transmitter, which we adapt from SISO CS-PTS \cite{b10}. Let $ \mathbf{\bar{X}}^{(i)} = \big[\bar{X}^{(i)} [0], \cdots, \bar{X}^{(i)} [N_c-1] \big]^T $ (for $ i = 1, \cdots, N_\mathrm{tx} $) denote the data symbols (e.g., PSK/QAM) at the $ i $-th antenna, where each has a duration of $ \Delta T $. Upon serial-to-parallel conversion, we obtain the symbols $ \mathbf{\widehat{X}}^{(i)} = \big[\widehat{X}^{(i)} [0], \cdots, \widehat{X}^{(i)} [N_c-1] \big]^T $ with duration $ N_c \Delta T $. Let $ \mathbf{H}_k \in \mathbb{C}^{N_\mathrm{rx} \times N_\mathrm{tx}} $ denote the channel between the transmitter and receiver in the $ k $-th subcarrier (for $ k = 1, \cdots, N_c $). Using SVD decomposition\footnote{We assume that the channel matrix is known. Therefore, SVD precoding and decoding can achieve the MIMO channel capacity \cite{telatar1999:capacity-multiantenna-gaussian-channels}. As a result, one data stream per singular value can be transmitted without causing interference.}, the channel matrix can be factorized into $\mathbf{H}_k = \mathbf{U}_k \mathbf{\Sigma}_k \mathbf{V}^{\dagger}_k $, where $ \dagger $ is the Hermitian transpose and $ \mathbf{U}_k \in \mathbb{C}^{N_\mathrm{rx} \times N_\mathrm{rx}} $, $ \mathbf{\Sigma}_k \in \mathbb{C}^{N_\mathrm{rx} \times N_\mathrm{tx}} $, $ \mathbf{V}_k \in \mathbb{C}^{N_\mathrm{tx} \times N_\mathrm{tx}} $. The right-hand matrix $ \mathbf{V}_k $ is used for precoding (at the transmitter) whereas $ \mathbf{U}_k $ is used for decoding (at the receiver). Thus, the precoded symbols at the $ k $-th subcarrier are computed as $ \mathbf{\breve{X}}_k = \mathbf{V}_k \mathbf{\widehat{X}}_k = \big[\breve{X}^{(1)} [k], \cdots, \breve{X}^{(N_\mathrm{tx})} [k] \big]^T $. Upon performing precoding on all subcarriers, we define $ \mathbf{\breve{X}}^{(i)} = \big[\breve{X}^{(i)} [0], \breve{X}^{(i)} [1],\cdots, \breve{X}^{(i)} [N_c-1] \big]^T $ at each antenna $ i $. Each $ \mathbf{\breve{X}}^{(i)} $ is split into $ M $ disjoint partitions, such that $ \mathbf{\breve{X}}^{(i)} = \sum_{m=1}^{M-1} \mathbf{\breve{X}}^{(i)}_m $. The IFFT is applied to every $ \mathbf{\breve{X}}^{(i)} $, thus yielding $ M $ partial transmit sequences\footnote{This name originates from prior literature, e.g., \cite{b9, b10}. Essentially, the qualification \emph{partial} stems from the fact that each $ \mathbf{\breve{x}}^{(i)}_m $ is a partial OFDM symbol formed by only a subset of modulated subcarriers from the total set that constitute a complete OFDM symbol.} $ \mathbf{\breve{x}}^{(i)}_m $ (for $ m = 0, \cdots, M-1 $) at transmit antenna $ i $. To preserve the orthogonality of SVD decomposition, the same phase rotations and time shifts must be applied to every $ m $-th partition (across all the antennas). Thus, the optimization problem collapses to finding the optimal phase rotations $ \left\lbrace \gamma^\mathrm{opt}_m\right\rbrace_{m = 0}^{M-1} $ and time shifts $ \left\lbrace \delta^\mathrm{opt}_m \right\rbrace_{m = 0}^{M-1} $ that minimize the maximum PAPR across all the antennas as shown in (\ref{e1}). 
\begin{equation} \label{e1}
\resizebox{0.9\columnwidth}{!}{
\begin{minipage}{1\columnwidth}
$ \mathcal{P}:
\left[ \left\lbrace \gamma^\mathrm{opt}_m \right\rbrace_{m=0}^{M-1}, \left\lbrace \delta^\mathrm{opt}_m \right\rbrace_{m=0}^{M-1} \right] = 
\underset{ \gamma_m \in \mathcal{U}, \\ \delta_m \in \mathcal{D} }{\arg\min} ~ \underset{1 \leq i \leq N_\mathrm{tx}}{\max} \mathrm{PAPR}_i $ 
\end{minipage}
}
\end{equation}
where the PAPR at the $ i $-th antenna is computed as\\
\resizebox{0.9\columnwidth}{!}{
\begin{minipage}{1\columnwidth}
$ \mathrm{PAPR}_i = \frac{\displaystyle \underset{0 \leq k \leq N_c L}{\max} \left| \sum_{m = 0}^{M - 1} \gamma_m \breve{x}^{(i)}_m \left[ k + \delta_m - N_c L \floor{\frac{k+\delta_m}{N_c L}} \right] \right|^2}
				{\displaystyle \frac{1}{N_c L} \sum_{k = 0}^{N_c L - 1} \left| \displaystyle\sum_{m = 0}^{M - 1} \gamma_m \breve{x}^{(i)}_m \left[ k + \delta_m - N_c L \floor{\frac{k+\delta_m}{N_c L}} \right] \right|^2 } $.
\end{minipage}
} \\ Each phase rotation $ \gamma_m $ is constrained to  the set $ \mathcal{U} = \left\lbrace e^{j\frac{2\pi}{U}u} \mid u = 0, \cdots, U-1 \right\rbrace $, where $ U $ represents the number of admissible phase rotations. Similarly, every time shift $ \delta_m $ is restricted to the set $ \mathcal{D} = \left\lbrace \frac{N_c}{D} d \mid d = 0, \cdots, D-1 \right\rbrace $, where $ D $ represents the number of possible time shifts. The signal to be transmitted via the $ i $-th antenna (prior to appending the cyclic prefix) is $ x^{(i)}[k] = \displaystyle\sum_{m=0}^{M-1} \gamma^\mathrm{opt}_m \breve{x}^{(i)}_m \Bigg[ k + \delta^\mathrm{opt}_m - N_c L \floor{\frac{k + \delta^\mathrm{opt}_m}{N_c L}} \Bigg] $ (for $ k = 1, \cdots, N_c $), where $ L $ is the oversampling factor. Due to time-frequency duality, cyclic time-domain shifting produces linear variation in the phase response. By cyclically delaying $ \mathbf{\breve{x}}_m $, phase variation per subcarrier can be achieved. Thus, every subcarrier $ k $ in the same $ m $-th partition will be rotated by an additional phase rotation $ \theta_m^{(k)} = \frac{2\pi k}{N_c}\delta_m $. The combined effect of both phase rotations and time shifts at subcarrier $ k $  of the $ m $-th partition is $ \gamma_m + \theta_m^{(k)} $.

\section{The Proposed \texttt{SWAN}} \label{s3}
\texttt{CSA} is inspired by the reproduction strategy of some cuckoo bird species that engage in brood parasitism to ensure their survival \cite{b15}. These birds deceive other species (host birds) by laying their eggs in their nests. This tactic relieves cuckoo birds from offspring feeding. As a result, more time can be devoted to food foraging and reproduction, thus improving the chances of survival of the species \cite{b15}. Sometimes, host birds are able to identify the cuckoo eggs and either abandon the nest or eject the parasite eggs. 

\noindent{\emph{Features}:} \texttt{CSA} captures the core reproduction strategy of cuckoo birds, which is succinctly described in the following.
\begin{itemize}[noitemsep, topsep = 0pt, leftmargin = 0.4cm]
	\item The initial population of $ N $ cuckoo birds is equal to the number of host nests.
	\item Each nest is a potential solution, and the suitability of each is defined by its fitness value. 
	\item The terms \emph{egg} and \emph{nest} are used interchangeably.
	\item The nests with the highest quality (i.e., highest fitness) will carry over the next generation of birds. 
	\item Host birds discover the parasite eggs with a probability $ p_a $.
\end{itemize}

\noindent{\emph{Drawbacks}:} 
\texttt{CSA} has an affordable computational complexity and remarkable exploration capability. However, \texttt{CSA} does not exploit the known solutions properly. By balancing exploration and exploitation, the search performance can be improved thereby attaining faster convergence. To achieve this balance, we integrate four low-complexity mechanisms, thus resulting into \texttt{SWAN}. Although \texttt{SWAN} reckons with additional features, the complexity remains affordable since the proposed improvements are applied to only a limited number of potential solutions. Algorithm \ref{a1} describes \texttt{SWAN} in detail. The devised mechanisms are described in the following.

\begin{algorithm}[!t]
\begin{algorithmic}[1]
		\footnotesize
		\State $ N $ : number of host nest: initial population of cuckoo birds 
		\State $ p_a $ : fraction of the total nests that represents the worst solutions
		\State $ n $ : number of dimensions
		\State $ \tau $ : counter of evaluations
		\State $ N_G $ : number of nests in the neighborhood of the best nest 		
		\State $ N_H $ : number of nests to be mutated 
			
		\Begin
			
			\State Define the objective function $ f:\mathbb{X}^n \to \mathbb{R} $  
		    \State Generate $ N $ host nests in $ \Omega = \{ \mathbf{z}_i \in \mathbb{X}^n \mid i = 1, \cdots, N \} $
			\State Evaluate the fitness $ F_{\mathbf{z}_i} $ of each nest $ \mathbf{z}_i $ 
			
			\While{ $ \sim \mathrm{stop criterion} $ }
				\State Find the current fittest nest $ \mathbf{z}_i $  in $ \Omega $ 
				\State Choose randomly another nest $ \mathbf{z}_i $ from $ \Omega $ avoiding $ \mathbf{z}_{\mathrm{best}} $ 
				\State Generate a new nest $ \mathbf{z}_j $ via a Levy flight from $ \mathbf{z}_i $  
				\State Evaluate the fitness $ F_{\mathbf{z}_j} $
				
				\If {($ F_{\mathbf{z}_j} > F_{\mathbf{z}_i} $)}
					\State Replace $ \mathbf{z}_i $ by the new solution $ \mathbf{z}_j $ 
				\EndIf
				
				\State Generate a new nest $ \mathbf{z}_k $ via a Levy flight from $ \mathbf{z}_{\mathrm{best}} $
				\State Evaluate the fitness $ F_{\mathbf{z}_k} $  
				
				\If {($ F_{\mathbf{z}_k} > F_{\mathrm{best}} $)}
					\State Replace $ \mathbf{z}_{\mathrm{best}} $  by the new solution $ \mathbf{z}_k $
				\EndIf
				        
				\State Update the counter $ \tau $ 
				
				\State \parbox[t]{\dimexpr\columnwidth-\leftmargin-\labelsep-\labelwidth-5pt}{ Let $ \Omega_Q = \{\mathbf{q}_1, \mathbf{q}_2, \mathbf{q}_3 \} $ be defined by the three nests with the highest fitness in $ \Omega $, such that 
				\begin{align} \nonumber
					\begin{cases} 
					\Omega_Q \subset \Omega, 
					\Omega_P \subset \Omega, 
					\Omega_Q \cup \Omega_P = \Omega, \\
					\Omega_Q \cap \Omega_P = \emptyset, 
					\Omega_P =\{ \mathbf{p}_i \in \mathbb{X}^n \mid i=1, \cdots, N-3\}.
					\end{cases}
				\end{align} \strut}
				\vspace*{-0.4cm} 
				\State Let the triangular region $ \mathcal{H} $ be defined by $ \mathbf{q}_1$, $ \mathbf{q}_2 $ and  $ \mathbf{q}_3 $	 
				\State Compute the parameters $ \mathbf{\Gamma} $ and $ \varepsilon $ of $ \mathcal{H} $
				\State Compute the parameter $ \mathbf{\Gamma}_{*} $
				\State \parbox[t]{\dimexpr\columnwidth-\leftmargin-\labelsep-\labelwidth-5pt}{ Define $ \Omega_R = \{\mathbf{r}_1, \mathbf{r}_2, \mathbf{r}_3 \} $ containing three potentially fitter nests obtained via Gaussian random walks using $ \mathbf{\Gamma}_{*} $ as a reference \strut}
				\State \parbox[t]{\dimexpr\columnwidth-\leftmargin-\labelsep-\labelwidth-5pt}{ Define $ \Omega_{\text{S}} = \Omega_Q \cup \Omega_R $ and sort the elements, $ F_{\mathbf{s}_i} \geq F_{\mathbf{s}_{i+1}} $ \strut}  
				\State \parbox[t]{\dimexpr\columnwidth-\leftmargin-\labelsep-\labelwidth-5pt}{ Replace $ \Omega_Q $ by the first 3 elements of $ \Omega_S $ \strut}
				\State Update the set $ \Omega $, such that $ \Omega = \Omega_P \cup \Omega_Q $
				\State Update the counter $ \tau $  
				\State \parbox[t]{\dimexpr\columnwidth-\leftmargin-\labelsep-\labelwidth-5pt}{ Build the subset $\Omega_A = \{ \mathbf{a}_i \in \mathbb{X}^n \mid i = 1, \cdots, N_A \} $ consisting of potentially $ N_A $ worst nests in $ \Omega $, such that 
				\begin{align} \nonumber
				\begin{cases} 
				\Omega_A \subset \Omega, \Omega_B \subset \Omega, 
				\Omega_A \cup \Omega_B = \Omega,
				\Omega_A \cap \Omega_B = \emptyset, \\
				\Omega_B = \{ \mathbf{b}_i \in \mathbb{X}^n \mid i = 1, \cdots, N_B \}, 
				N_A = p_a N
				\end{cases}
				\end{align} \strut}
				\vspace*{-0.4cm} 
				\State \parbox[t]{\dimexpr\columnwidth-\leftmargin-\labelsep-\labelwidth-5pt}{ Build a subset $ \Omega_C = \{\mathbf{c}_i \in \mathbb{X}^n \mid i = 1, \cdots,N_A \} $ consisting of $ N_A $ potentially fitter nests than those contained in $ \Omega_A $ \strut}
				
				\State Build the set $ \Omega_D = \Omega_A \cup \Omega_C $ consisting of $ 2N_A $ elements 
				
				\State \parbox[t]{\dimexpr\columnwidth-\leftmargin-\labelsep-\labelwidth-5pt}{ Sort $ \Omega_D = \{ \mathbf{d}_i \in \mathbb{X}^n \mid i = 1, \cdots, 2N_A \} $, with $ F_{\mathbf{d}_i} \geq F_{\mathbf{d}_{i+1}} $ \strut}
				
				\State \parbox[t]{\dimexpr\columnwidth-\leftmargin-\labelsep-\labelwidth-5pt}{ Replace the elements of $ \Omega_A $  by the first $ N_A $ elements of $ \Omega_D $, such that $ \Omega_A = \{\mathbf{d}_i \in \mathbb{X}^n \mid i = 1, \cdots ,N_A \} $ \strut}
				
				\State Update the set $ \Omega $, such that $ \Omega_A \cup \Omega_B = \Omega $
				
				\State Update the counter $ \tau $  
				
				\State Find the current best nest $ \mathbf{z}_{\mathrm{best}} $  
 
				\State \parbox[t]{\dimexpr\columnwidth-\leftmargin-\labelsep-\labelwidth-5pt}{ Build $ \Omega_G = \{\mathbf{g}_i \in \mathbb{X}^n \mid i = 1, \cdots, N_G \}$ with $ N_G $ new nests generated in the neighborhood of $ \mathbf{z}_{\mathrm{best}} $ using uniform random walks\strut}
				
				\State \parbox[t]{\dimexpr\columnwidth-\leftmargin-\labelsep-\labelwidth-5pt}{ Select randomly $ N_G $ nests from $ \Omega $ avoiding $ \mathbf{z}_{\mathrm{best}} $ and replace them by the newly generated eggs in $ \Omega_G $ \strut}
				
				\State Update the counter $ \tau $ 
				
				\State \parbox[t]{\dimexpr\columnwidth-\leftmargin-\labelsep-\labelwidth-5pt}{ Build the set $ \Omega_H = \{\mathbf{h}_i \in \mathbb{X}^n \mid i = 1, \cdots, N_H \}$ containing the $ \Omega_H $ least fit nests and apply mutation \strut}
				
				\State Keep the current best solutions for the next generation 
				
				\State Update the counter $ \tau $  
				        	
			\EndWhile
			
		\End
		\caption{\texttt{SWAN} algorithm for PAPR reduction} 
		\label{a1}
	\end{algorithmic}
\end{algorithm}

\noindent{\textbf{Mechanism 1} \textit{(Update of the best solution)}:} This mechanism is explained in lines $ 20 \sim 24 $ of Algorithm \ref{a1}. It deals with the appropriate placement of the best nest in the search space. In \texttt{CSA}, the nest with the highest fitness in each generation is used as a reference for generating Levy flights for the rest of the nests. However, the best nest is not updated until the next iteration (or generation). Thus, \texttt{SWAN} replaces the best nest with a more suitable one if a higher fitness is obtained. Specifically, the nests are generated according to
\begin{equation} \label{e2}
\resizebox{0.91\columnwidth}{!}{
\begin{minipage}{0.95\columnwidth}
$
	\mathbf{z}^{(t+1)}_i= 
	\begin{cases}
	\mathbf{z}^{(t)}_i + \alpha \Big(\mathbf{z}^{(t)}_{\mathrm{best}} - \mathbf{z}^{(t)}_i \Big) \odot \mathbf{w}_{\mathrm{lev}} &\mbox{if } \mathbf{z}^{(t)}_i \neq \mathbf{z}^{(t)}_{\mathrm{best}} \\
	\mathbf{z}^{(t)}_i + \alpha \mathbf{w}_{\mathrm{lev}} &\mbox{if } \mathbf{z}^{(t)}_i = \mathbf{z}^{(t)}_{\mathrm{best}},
	\end{cases}  
$
\end{minipage}
}	
\end{equation}
where $ \mathbf{z}^{(t)}_{\mathrm{best}} = \big[z^{(t)}_{\mathrm{best}}[1], \cdots, z^{(t)}_{\mathrm{best}}[n] \big]^T $ is the best solution at iteration $ t $, which is used as a reference for deriving new candidate solutions $ \mathbf{z}^{(t+1)}_i = \big[ z^{(t+1)}_i[1], \cdots,z^{(t+1)}_i[n] \big]^T $ (for $ i = 1, \cdots, N $). The random walks $ \mathbf{w}_{\mathrm{lev}} = \big[ w_{\mathrm{lev}}[1], \cdots,w_{\mathrm{lev}}[n] \big]^T $ are drawn from a Levy distribution function \cite{b15}. In (\ref{e2}), $ \alpha $ is a scaling factor, $ n $ denotes the dimensions of the solution, and $ \odot $ represents element-wise multiplication. 

\noindent{\textbf{Mechanism 2} \textit{(Best triad mating)}:}
In \texttt{CSA}, birds display limited social interaction. However, \texttt{SWAN} fosters collaborative information sharing, which improves convergence. We introduce the idea of \emph{best triad mating}, which exploits information available at the best three solutions, intending to intensify the search in a smaller space within which (with high probability) a better solution may lie. The procedure consists of five steps that have been summarized in lines $ 26 \sim 34 $ of Algorithm \ref{a1}.

\begin{figure}[t!]
	\centering
	\includegraphics[width=0.65\columnwidth]{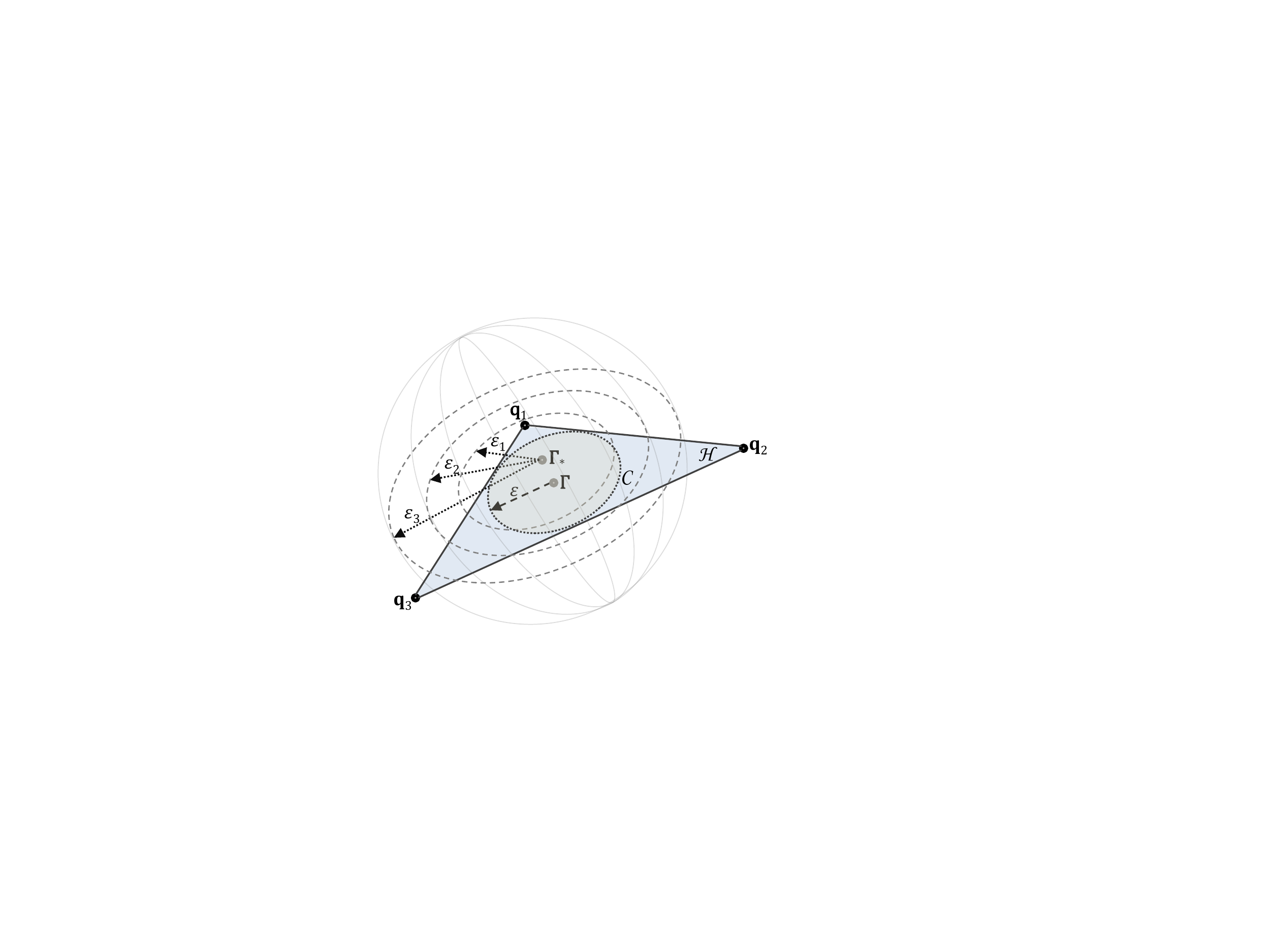}
	\caption{Best triad mating process (Mechanism 2)}
	\vspace{-2mm}
	\label{f2}
	\vspace{-3mm}
\end{figure}

\noindent{\emph{Step 2.1:}} 
Let $ \Omega_Q = \{\mathbf{q}_1, \mathbf{q}_2, \mathbf{q}_3 \} $ be a subset of $ \Omega $ containing the fittest three nests (sorted in descending order of their fitness values). Further, let $ \Omega_P $ be a subset of $ \Omega $ representing the complement of $ \Omega_Q $. Using the elements in $ \Omega_Q $, we define $ \mathcal{H} $ (with point-to-point distances $ \| \mathbf{q}_1 - \mathbf{q}_2 \|, \| \mathbf{q}_2 - \mathbf{q}_3 \|, \| \mathbf{q}_3 - \mathbf{q}_1 \| $) as shown in Fig. \ref{f2}. These three solutions $ \{\mathbf{q}_1, \mathbf{q}_2, \mathbf{q}_3 \} $ achieve the highest fitness in the generation $ t $. However, potentially fitter solutions might lie in a neighboring area to them. 

\noindent{\emph{Step 2.2:}} 
We calculate $ \mathbf{\Gamma} $ and $ \varepsilon $ (shown in Fig. \ref{f2}) as follows

\resizebox{0.85\columnwidth}{!}{
\begin{minipage}{1.05\columnwidth}
\begin{equation} \nonumber
	\mathbf{\Gamma} = \frac{\displaystyle\sum^{2}_{i=0} \bigg\|\mathbf{q}_{i-b} - \mathbf{q}_{i-c} \bigg\|_2 \mathbf{q}_{i-a}} 
	{ \displaystyle\sum^{2}_{i=0} \bigg\| \mathbf{q}_{i-a} - \mathbf{q}_{i-b} \bigg\|_2 } ~~ 
	\varepsilon = \sqrt{8\displaystyle\prod_{i=0}^{2} \left\{ \frac{1}{2} - \frac{\bigg\|\mathbf{q}_{i-b} - \mathbf{q}_{i-c} \bigg\|_2}{\displaystyle\sum^{2}_{i=0} \bigg\| \mathbf{q}_{i-a} - \mathbf{q}_{i-b} \bigg\|_2} \right\}}
\end{equation}
\end{minipage}
}

\noindent where $ a = 3\smallfloor{\frac{i}{3}} + 1 $, $ b = 3\smallfloor{\frac{i + 1}{3}} + 2 $, $ c = 3\smallfloor{\frac{i + 2}{3}} + 3 $. Essentially, $ \mathbf{q}_1 $, $ \mathbf{q}_2 $ and $ \mathbf{q}_3 $ delimit a triangular region $ \mathcal{H} $ with sides $ \| \mathbf{q}_1 - \mathbf{q}_2 \|, \| \mathbf{q}_2 - \mathbf{q}_3 \|, \| \mathbf{q}_3 - \mathbf{q}_1 \| $. Thus, $ \mathbf{\Gamma} $ is the in-center of the circle $ \mathcal{C} $ inscribed in $ \mathcal{H} $ whereas $ \varepsilon $ is the in-radius of $ \mathcal{C} $.

\noindent{\emph{Step 2.3:}}  
Let $ F_{\mathbf{q}_1} $, $ F_{\mathbf{q}_2} $, $ F_{\mathbf{q}_3} $ be the fitness values of $ \mathbf{q}_1 $, $ \mathbf{q}_2 $, $ \mathbf{q}_3 $, respectively. We compute the weighted reference $ \mathbf{\Gamma}_{*} $ via
\begin{equation} \label{e5}
	\mathbf{\Gamma}_{*} = \frac{F_{\mathbf{q}_2} + F_{\mathbf{q}_3}}{F_{\mathbf{q}_1} + F_{\mathbf{q}_2} + F_{\mathbf{q}_3}}\mathbf{\Gamma} + \frac{F_{\mathbf{q}_1}}{F_{\mathbf{q}_1} + F_{\mathbf{q}_2} + F_{\mathbf{q}_3}}\mathbf{q}_1.
\end{equation}

When $ F_1 \ge F_2 + F_3 $, $\mathbf{q}_1$ has a weight higher than $\mathbf{\Gamma}$. This indicates a higher fitness of $ \mathbf{q}_1 $ compared to the other two solutions. Thus, $ \mathbf{\Gamma}_{*} $ will lean towards $\mathbf{q}_1$.

\noindent{\emph{Step 2.4:}} 
Three new solutions $ \Omega_R = \{\mathbf{r}_1, \mathbf{r}_2, \mathbf{r}_3\} $ are generated using $ n $-dimensional Gaussian random walks $ \mathbf{w}_{\mathrm{gau}} = [ w_{\mathrm{gau}}[1], \cdots,w_{\mathrm{gau}}[n] ]^T $ by means of 
\begin{equation} \label{e6}
	\begin{cases}
	\mathbf{r}^{(0)}_\ell=\mathbf{\Gamma}_{*}\\
	\mathbf{r}^{(t+1)}_\ell=\mathbf{r}^{(t)}_\ell + \kappa_{\ell} \mathbf{w}_{\mathrm{gau}} &\mbox{if } t > 0,
	\end{cases}
\end{equation}
where $ \kappa_\ell = \varepsilon_{\ell} v + \varepsilon_{\ell} \phi \big ( \max\big\{\frac{F_1-(F_2+F_3)}{F_1+F_2+F_3}, 0\big\} \big ) $ and $ \varepsilon_{\ell} = \ell \frac{\varepsilon}{2} $, (for $ \ell = 1, 2, 3 $) . Suitable values for $ \phi $ and $ \nu $ are in the ranges $ 0.10 \leq \phi \leq 0.45 $, $ 0.55 \leq \nu \leq 0.90 $, which have been obtained via Monte Carlo simulation with standard benchmark functions: hyperdimensional sphere \cite{b17}, Ackley \cite{b18}, Michalewicz \cite{b19}, Griewank \cite{b20}, and Easom \cite{b21}. By generating new $ \{ \mathbf{r}_1, \mathbf{r}_2, \mathbf{r}_3 \} $ in the proximity of $ \mathbf{\Gamma}_{*} $, the search is confined to a smaller but potentially richer space, thereby improving convergence.  

\noindent{\emph{Step 2.5:}}
Let $ \Omega_S $ be defined as $\Omega_S = \Omega_Q \cup \Omega_R $, thus consisting of $ \{ \mathbf{q}_1, \mathbf{q}_2, \mathbf{q}_3 \} $ and the newly generated $ \{ \mathbf{r}_1, \mathbf{r}_2, \mathbf{r}_3 \} $. Let the elements of $ \Omega_S $ be sorted in descending order of fitness, such that  $ F_{\mathbf{s}_i} \geq F_{\mathbf{s}_{i+1}} $ for all $ \mathbf{s}_i \in \Omega_S $. Now, we redefine $ \Omega_Q $ such that it contains the three best solutions of $ \Omega_S $, i.e. $ \Omega_Q = \{ \mathbf{s}_1, \mathbf{s}_2, \mathbf{s}_3 \} $. Finally, we let $ \Omega $ be the union of $ \Omega_Q $ and $ \Omega_P $ (defined in \emph{Step 2.1}), i.e., $ \Omega = \Omega_Q \cup \Omega_P $ (note that the cardinality of $ \Omega $ has not changed).
 
\noindent{\textbf{Mechanism 3} \textit{(Exploitation of the best nest)}:}
While each nest in \texttt{CSA} accommodates only one egg, we allow \texttt{SWAN} to accommodate more than one egg per nest as described in lines $ 42 \sim 44 $ of Algorithm \ref{a1}. Specifically, this mechanism intensifies the exploitation of the best known solution as follows. A random integer $ N_G = \left\lbrace 0, 1, 2, 3 \right\rbrace $ is drawn with equal probability. $ N_G $ is the number of solutions randomly selected from $ \Omega $ which are to be replaced by new solutions $ \Omega_G = \{ \mathbf{g}_1, \cdots, \mathbf{g}_{N_G} \} $.
By means of $n$-dimensional uniform random walks $ \mathbf{w}_{\mathrm{uni}} = [ w_{\mathrm{uni}}[1], \cdots, w_{\mathrm{uni}}[n] ]^T $, additional $ N_G $ solutions are generated as shown in (\ref{e9}) (for $ \ell = 1, \cdots, N_G $), where $ \mathbf{\Psi}_1 = [ \psi_1[1], \cdots, \psi_1[n] ]^T $ is a vector whose elements are obtained from a normalized Gaussian probability density function.
\begin{equation} \label{e9}
	\begin{cases}
		\mathbf{g}^{(0)}_\ell=\mathbf{z}_{\mathrm{best}}\\
		\mathbf{g}^{(t+1)}_\ell=\mathbf{g}^{(t)}_\ell + \big( \mathbf{g}^{(0)} + 0.25\mathbf{\Psi}_1 \big) \odot \mathbf{w}_{\mathrm{uni}} & \mbox{if } t > 0.
	\end{cases}
\end{equation} 
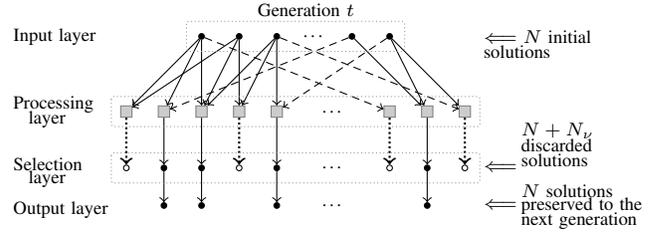
\begin{figure}[!t]
	\centering
	\begin{tikzpicture}[inner sep=0.75mm]
	\tikzstyle{every node} = [font=\scriptsize]
	\node (a1) at (0,0) [circle, draw, inner sep = 0.75, fill = black] {};
	\node (a2) at (0.5,0) [circle, draw, inner sep = 0.75, fill = black] {};
	\node (a3) at (1,0) [circle, draw, inner sep = 0.75, fill = black] {};
	\node at (1.5,0) {\dots};
	\node (a4) at (2,0) [circle,draw, inner sep=0.75,fill=black] {};
	\node (a5) at (2.5,0) [circle,draw, inner sep=0.75,fill=black] {};
	
	\node (b1) at (-1,-1) [rectangle, draw = black!50, fill = black!20] {};
	\node (b2) at (-0.5,-1) [rectangle, draw=black!50, fill = black!20] {};
	\node (b3) at (0,-1) [rectangle, draw=black!50, fill = black!20] {};
	\node (b4) at (0.5,-1) [rectangle, draw=black!50, fill = black!20] {};
	\node (b5) at (1,-1) [rectangle, draw=black!50, fill = black!20] {};	
	\node at (1.75,-1) {\dots};
	\node (b6) at (2.5,-1) [rectangle,draw=black!50,fill=black!20] {};
	\node (b7) at (3,-1) [rectangle,draw=black!50,fill=black!20] {};
	\node (b8) at (3.5,-1) [rectangle,draw=black!50,fill=black!20] {};

	\node (c1) at (-0.5,-1.75) [circle, draw, inner sep=0.75,fill=black] {};
	\node (c2) at (0,-1.75) [circle,draw, inner sep=0.75,fill=black] {};
	\node (c3) at (1,-1.75) [circle,draw, inner sep=0.75,fill=black] {};
	\node at (1.75,-2.25) {\dots};	
	\node (c4) at (3,-1.75) [circle,draw, inner sep=0.75,fill=black] {};
	
	\node (m1) at (-0.5,-2.25) [circle,draw, inner sep=0.75,fill=black] {};
	\node (m2) at (0,-2.25) [circle,draw, inner sep=0.75,fill=black] {};
	\node (m3) at (1,-2.25) [circle,draw, inner sep=0.75,fill=black] {};
	\node (m4) at (3,-2.25) [circle,draw, inner sep=0.75,fill=black] {};	
	
	\node (x1) at (-1,-1.75) [circle,draw, inner sep=0.75,fill=white]{};
	\node (x2) at (0.5,-1.75) [circle,draw, inner sep=0.75,fill=white]{};
	\node (x3) at (2.5,-1.75) [circle,draw, inner sep=0.75,fill=white]{};
	\node (x4) at (3.5,-1.75) [circle,draw, inner sep=0.75,fill=white]{};
	\node at (1.75,-1.75) {\dots};	
	
	\draw [->] (a1) -- (b1);
	\draw [->] (a2) -- (b1);
	
	\draw [->] (a1) -- (b2);
	\draw [densely dashed, ->] (a4) -- (b2);
	
	\draw [->] (a1) -- (b3);
	\draw [->] (a2) -- (b3);
	\draw [->] (a3) -- (b3);
	
	\draw [->] (a2) -- (b4);
	\draw [->] (a3) -- (b4);
	
	\draw [->] (a3) -- (b5);
	\draw [densely dashed, ->] (a5) -- (b5);
	
	\draw [densely dashed, ->] (a1) -- (b6);
	
	\draw [->] (a4) -- (b7);
	\draw [->] (a5) -- (b7);
	
	\draw [densely dashed, ->] (a3) -- (b8);
	\draw [->] (a5) -- (b8);
	
	\draw [thick, densely dotted, ->] (b1) -- (x1);
	\draw [thick, densely dotted, ->] (b4) -- (x2);
	\draw [thick, densely dotted, ->] (b6) -- (x3);
	\draw [thick, densely dotted, ->] (b8) -- (x4);
	
	\draw [->] (b2) -- (c1);
	\draw [->] (b3) -- (c2);
	\draw [->] (b5) -- (c3);
	\draw [->] (b7) -- (c4);
	
	\draw [->] (c1) -- (m1);
	\draw [->] (c2) -- (m2);
	\draw [->] (c3) -- (m3);
	\draw [->] (c4) -- (m4);
	
	\node[text width=3.5cm] at (5.5,0) {$\Longleftarrow N$ initial};
	\node[text width=3.5cm] at (5.5,-0.2) {solutions};
	
	\node[text width=3.5cm] at (5.5,-1.75) {$\Longleftarrow$};
	\node[text width=3.5cm] at (6,-1.25) {$N+N_{\nu}$};
	\node[text width=3.5cm] at (6,-1.45) {discarded};
	\node[text width=3.5cm] at (6,-1.65) {solutions};
	
	\node[text width=3.5cm] at (5.5,-2.25) {$\Longleftarrow$};
	\node[text width=3.5cm] at (6,-2.05) {$N$ solutions };
	\node[text width=3.5cm] at (6,-2.25) {preserved to the};
	\node[text width=3.5cm] at (6,-2.45) {next generation};
	
	\node[text width=3cm] at (-1,0) {Input layer};
	\node[text width=3cm] at (-1,-0.9) {Processing};
	\node[text width=3cm] at (-0.8,-1.1) {layer};
	\node[text width=3cm] at (-1,-2.3) {Output layer};
	\node[text width=3cm] at (-1,-1.7) {Selection};
	\node[text width=3cm] at (-0.8,-1.9) {layer};
	
	\node[text width=3.5cm] at (2.5,0.35) {Generation $t$};
	
	\draw[densely dotted,rounded corners=1,color=gray!80]($(a1)+(-.2,.2)$)rectangle($(a5)+(.2,-.2)$);
	\draw[densely dotted,rounded corners=1,color=gray!80]($(b1)+(-.2,.2)$)rectangle($(b8)+(.2,-.2)$);
	\draw[densely dotted,rounded corners=1,color=gray!80]($(x1)+(-.2,.2)$)rectangle($(x4)+(.2,-.2)$);
	
	\end{tikzpicture}
	\vspace{-3mm}
	\caption{Generalized structure of \texttt{SWAN}}
	\label{f3}
	\vspace{-5mm}
\end{figure}
\begin{figure} [!t]
	\centering
	\begin{tikzpicture}[inner sep=0.75mm]
	\tikzstyle{every node} = [font=\scriptsize]
	\node (x1) at (-0.5,0.7) [circle,draw, inner sep=0.75, fill=black] {};
	\node (x2) at (0.5,0.7) [circle,draw, inner sep=0.75,fill=black] {};
	\node (x3) at (1.5,0.7) [circle,draw, inner sep=0.75,fill=black] {};
	\node (x4) at (2.5,0.7) [circle,draw, inner sep=0.75,fill=black] {};
	\node (x5) at (3.5,0.7) [circle,draw, inner sep=0.75,fill=black] {};
	\node at (4.15,0.7) {\dots};
	\node (x6) at (4.75,0.7) [circle,draw, inner sep=0.75,fill=black] {};	
	\node (x7) at (5.75,0.7) [circle,draw, inner sep=0.75,fill=black] {};	
	
	\node (a1) at (0,0.1) [circle split,draw, double, inner sep=1.6] {};
	\node (b1) at (1.5,0.1) [circle split,draw, double, inner sep=1.6] {};
	\node (c1) at (3,0.1) [circle split,draw, double, inner sep=1.6] {};
	\node at (4.15,0.1) {\dots};
	\node (d1) at (5.25,0.1) [circle split,draw, double, inner sep=1.6] {};
	
	\node (a2) at (-0.5,-0.6) [circle,draw, inner sep=0.75,fill=black] {};
	\node (a3) at (-0.25,-0.6) [circle,draw, inner sep=0.75,fill=black] {};
	\node at (0.15,-0.6) {\dots};
	\node (a4) at (0.5,-0.6) [circle,draw, inner sep=0.75,fill=black] {};

	\node (b2) at (1,-0.6) [circle,draw, inner sep=0.75,fill=black] {};
	\node (b3) at (1.25,-0.6) [circle,draw, inner sep=0.75,fill=black] {};
	\node at (1.65,-0.6) {\dots};
	\node (b4) at (2,-0.6) [circle,draw, inner sep=0.75,fill=black] {};
	
	\node (c2) at (2.5,-0.6) [circle,draw, inner sep=0.75,fill=black] {};
	\node (c3) at (2.75,-0.6) [circle,draw, inner sep=0.75,fill=black] {};
	\node at (3.15,-0.6) {\dots};
	\node (c4) at (3.5,-0.6) [circle,draw, inner sep=0.75,fill=black] {};

	\node at (4.15,-0.6) {\dots};
	\node (d2) at (4.75,-0.6) [circle,draw, inner sep=0.75,fill=black] {};
	\node (d3) at (5,-0.6) [circle,draw, inner sep=0.75,fill=black] {};
	\node at (5.4,-0.6) {\dots};
	\node (d4) at (5.75,-0.6) [circle,draw, inner sep=0.75,fill=black] {};
	
	\draw [->] (a1) -- (a2);
	\draw [->] (a1) -- (a3);
	\draw [->] (a1) -- (a4);

	\draw [->] (b1) -- (b2);
	\draw [->] (b1) -- (b3);
	\draw [->] (b1) -- (b4);
	
	\draw [->] (c1) -- (c2);
	\draw [->] (c1) -- (c3);
	\draw [->] (c1) -- (c4);
	
	\draw [->] (d1) -- (d2);
	\draw [->] (d1) -- (d3);
	\draw [->] (d1) -- (d4);
	
	\draw [->] (x1) -- (-0.5,0.3);
	\draw [->] (x2) -- (0.5,0.3);
	\draw [->] (x3) -- (1.5,0.3);
	\draw [->] (x4) -- (2.5,0.3);
	\draw [->] (x5) -- (3.5,0.3);
	\draw [->] (x6) -- (4.75,0.3);
	\draw [->] (x7) -- (5.75,0.3);
	
	\draw [->] (a2) -- (-0.5,-1);
	\draw [->] (a3) -- (-0.25,-1);
	\draw [->] (a4) -- (0.5,-1);
	\node at (0.15,-1) {\dots};
	
 	\draw [->] (b2) -- (1,-1);
 	\draw [->] (b3) -- (1.25,-1);
 	\draw [->] (b4) -- (2,-1);
 	\node at (1.65,-1) {\dots};
 	
 	\draw [->] (c2) -- (2.5,-1);
 	\draw [->] (c3) -- (2.75,-1);
 	\draw [->] (c4) -- (3.5,-1);
 	\node at (3.15,-1) {\dots}; 	

	\node at (4.15,-1) {\dots};
 	\draw [->] (d2) -- (4.75,-1);
 	\draw [->] (d3) -- (5,-1);
 	\draw [->] (d4) -- (5.75,-1);
 	\node at (5.4,-1) {\dots}; 	
	
	\draw[decoration={brace,mirror,raise=5pt},decorate] (5.8,0.7) -- node[above=8pt] {$N$ nests} (-0.55,0.7);
	\draw[decoration={brace,mirror,raise=5pt},decorate] (-0.6,-0.9) -- node[below=8pt] {$\rho_2$ nests} (0.6,-0.9);
	\draw[decoration={brace,mirror,raise=5pt},decorate] (0.9,-0.9) -- node[below=8pt] {$\rho_2$ nests} (2.1,-0.9);
	\draw[decoration={brace,mirror,raise=5pt},decorate] (2.4,-0.9) -- node[below=8pt] {$\rho_2$ nests} (3.6,-0.9);
	\draw[decoration={brace,mirror,raise=5pt},decorate] (4.65,-0.9) -- node[below=8pt] {$\rho_{2}$ nests} (5.85,-0.9);
	\draw[decoration={brace,mirror,raise=5pt},decorate] (-0.7,-1.3) -- node[below=8pt] {$\rho=\rho_1 \rho_2$ nests} (5.95,-1.3);
	\draw [decorate,decoration={brace,amplitude=4pt,aspect=0.25}] (5.4,-0.1) -- (-0.15,-0.1) node[below=8pt, pos=0.32, right] {$\rho_1$ nests} (5.95,-2.25);
	
	\draw[densely dotted,rounded corners=1]($(-0.75,-0.5)+(-.2,1)$)rectangle($(6,-0.5)+(.2,-.3)$);
	\draw[densely dashed,rounded corners=2]($(-0.5,0)+(-.2,.3)$)rectangle($(5.75,0)+(.2,-.1)$);
	
	\node[text width=3cm] at (-0.85,0.7) {Input layer};
	\node[text width=3cm] at (-0.85,-0.5) {Processing};
	\node[text width=3cm] at (-0.85,-0.7) {layer};
	\node[text width=3cm] at (-0.85,0.2) {Intermediate};
	\node[text width=3cm] at (-0.85,0) {node};

	\end{tikzpicture}
	\vspace{-3mm}
	\caption{Adaptation of \texttt{SWAN}}
	\vspace{-5mm}
	\label{f4}
\end{figure}

\noindent{\textbf{Mechanism 4} \textit{(Mutation of the worst nests)}:}
This mechanism, described in lines $ 46 \sim 48 $, creates new nests in different locations in order to replace only a subset of the worst-performing nests. This fosters balance between exploitation and exploration by means of regulating diversification of new solutions and re-usage of the old ones. Thus, we define $ \Omega_H = \{\mathbf{h}_1, \cdots, \mathbf{h}_{N_H} \} $ containing the least fit $ N_H $ solutions from $ \Omega $. Then, each element of $ \Omega_H $ is updated via (\ref{e10}) only if the fitness of the new solution $ \mathbf{h}^{(t+1)}_\ell $ (for $ \ell = 1, \cdots, N_H $) has increased with respect to that of the previous $ \mathbf{h}^{(t)}_\ell $
\begin{equation} \label{e10}
	\mathbf{h}^{(t+1)}_\ell=\frac{1}{2} \Big(\mathbf{h}^{(t)}_\ell + \mathbf{h}^{(t)}_{\mathrm{w}}\Big) \odot \mathbf{\Psi}_2 \odot \mathbf{\Psi}_3,
\end{equation}
where $ \mathbf{h}_{\mathrm{w}} = [ h_{\mathrm{w}}[1], \cdots, h_{\mathrm{w}}[n] ]^T $ represents the nests with the lowest fitness at iteration $ t $. The elements of $ \mathbf{\Psi}_2 = [ \psi_2[1], \cdots, \psi_2[n] ]^T $ are obtained from a random variable uniformly distributed in the range $ [1,2] $ whereas the elements of $ \mathbf{\Psi}_3 = [ \psi_3[1], \cdots, \psi_3[n] ]^T $ are 1 or -1 with equal probability. 

\emph{Remark:} \texttt{CSA} is initialized with a set of $ N $ nests (or solutions). However, within each generation, $ 2N $ solutions are generated from which only $ N $ are carried over to the next generation. \texttt{SWAN} is also initialized with $ N $ eggs and within a generation $ 2N + N_{\nu} $ eggs are generated. Nevertheless, only $ N $ are preserved as shown in Fig. \ref{f3}. The difference, $ N_{\nu} = 3 + N_G + N_H $, is due to the proposed mechanisms.
\begin{figure*}
\begin{minipage}{0.31\textwidth}
	\centering
	\usetikzlibrary{spy}
	\begin{tikzpicture}[scale=0.7, spy using outlines={circle, magnification=4, connect spies}]
	\begin{semilogyaxis}
	[
	height=6.5cm,
	width=8.4cm,
	xlabel={$\mathrm{PAPR}_0$ (dB)},
	ylabel={$P(\mathrm{PAPR} > \mathrm{PAPR}_0)$},
	xmin=4, xmax=12, grid=both,
	major grid style={black!40},
	ymin=0.0001, ymax=1.5,
	xtick={4, 5, 6, 7, 8, 9, 10, 11, 12},
	legend entries={SISO, SVD $ 2 \times 2 $ , SVD $ 3 \times 3 $, SVD $ 4 \times 4 $, SVD $ 6 \times 6 $},
	legend style={cells={anchor=west}, at={(0.03,0.03)}, anchor=south west, font=\scriptsize},
	]
	\addplot [mark=oplus*, mark options={scale=1,solid, fill=white},] table {./data/data_siso.dat};
	\addplot [densely dashed, mark=*, mark options={scale=1,solid},] table {./data/data_svd_2x2.dat};
	\addplot [mark=triangle*, mark options={scale=1,solid, fill=white},] table {./data/data_svd_3x3.dat};
	\addplot [mark=diamond*,] table {./data/data_svd_4x4.dat};
	\addplot [mark=pentagon*, mark options={scale=1,solid, fill=white},] table {./data/data_svd_6x6.dat};
	
	\coordinate (spypoint) at (8.7,0.000065);
	\coordinate (magnifyglass) at (5,0.005);
	\end{semilogyaxis}
	
	\spy [black, size=15mm] on (spypoint)
	in node[fill=white] at (magnifyglass);
	\end{tikzpicture}
	\vspace{-6mm}
	\caption{Effect of the number of transmit antennas on the PAPR}
	\label{f5}
\end{minipage}
\hspace{2mm}
\begin{minipage}{0.31\textwidth}
	\centering
	\begin{tikzpicture}[scale=0.7]
	\begin{semilogyaxis}
	[
	height=6.5cm,
	width=8.4cm,
	xlabel={$\mathrm{PAPR}_0$ (dB)},
	ylabel={$P(\mathrm{PAPR} > \mathrm{PAPR}_0)$},
	xmin=4, xmax=12, grid=both,
	major grid style={black!40},
	ymin=0.0001, ymax=1.2,
	xtick={4, 5, 6, 7, 8, 9, 10, 11, 12},
	xtick={4, 5, 6, 7, 8, 9, 10, 11, 12},
	legend entries={No optimization, $ V = 4 $, $ V = 16 $, $ V = 64 $, $ V = 256 $, $ V = 1024 $, $ V = 4096 $},
	legend style={cells={anchor=west}, at={(0.99,0.99)}, font=\scriptsize},
	]
	
	\addplot [dashdotdotted,] table {./data/data_svd_4x4.dat};
	\addplot [mark=triangle*, mark options={scale=1,solid},] table {./data/data_svd_4x4_slm_4.dat};
	\addplot [densely dashed, mark=diamond*, mark options={scale=1,solid,fill=white},] table {./data/data_svd_4x4_slm_16.dat};
	\addplot [mark=pentagon*, mark options={scale=1,solid},] table {./data/data_svd_4x4_slm_64.dat};
	\addplot [densely dashed, mark=*, mark options={scale=1,solid,fill=white},] table {./data/data_svd_4x4_slm_256.dat};
	\addplot [mark=square*, mark options={scale=1,solid},] table {./data/data_svd_4x4_slm_1024.dat};
	\addplot [densely dashed, mark=oplus*, mark options={scale=1,solid,fill=white},] table {./data/data_svd_4x4_slm_4096.dat};
	
	\end{semilogyaxis}
	\end{tikzpicture}
	\vspace{-6mm}
	\caption{PAPR reduction performance in SVD-MIMO $ 4 \times 4$ using SLM}
	\label{f6}
\end{minipage}
\hspace{2mm}
\begin{minipage}{0.31\textwidth}
	\centering
	\begin{tikzpicture}[scale=0.7]
	\begin{semilogyaxis}
	[
	height=6.5cm,
	width=8.4cm,
	xlabel={$\mathrm{PAPR}_0$ (dB)},
	ylabel={$P(\mathrm{PAPR} > \mathrm{PAPR}_0)$},
	xmin=4, xmax=12, grid=both,
	major grid style={black!40},
	ymin=0.0001, ymax=1.5,
	xtick={4, 5, 6, 7, 8, 9, 10, 11, 12},
	legend entries={No optimization, {$M = 2, U = 2$}, {$M = 2, U = 4$}, {$M = 2, U = 8$}, {$M = 4, U = 2$}, {$M = 4, U = 4$}, {$M = 8, U = 2$}, {$M = 8, U = 4$}},
	legend style={cells={anchor=west}, at={(0.99,0.99)}, font=\scriptsize},
	]
	
	\addplot [dashdotdotted,] table {./data/data_svd_4x4.dat};
	\addplot [mark=triangle*, mark options={scale=1,solid},] table {./data/data_svd_4x4_pts_2p_2w.dat};
	\addplot [densely dashed, mark=diamond*, mark options={scale=1,solid, fill=white},] table {./data/data_svd_4x4_pts_2p_4w.dat};
	\addplot [mark=pentagon*, mark options={scale=1,solid},] table {./data/data_svd_4x4_pts_2p_8w.dat};
	\addplot [densely dashed, mark=*, mark options={scale=1,solid,fill=white},] table {./data/data_svd_4x4_pts_4p_2w.dat};
	\addplot [mark=square*, mark options={scale=1,solid},] table {./data/data_svd_4x4_pts_4p_4w.dat};
	\addplot [densely dashed, mark=oplus*, mark options={scale=1,solid, fill=white},] table {./data/data_svd_4x4_pts_8p_2w.dat};
	\addplot [mark=10-pointed star, mark options={scale=1,solid},] table {./data/data_svd_4x4_pts_8p_4w.dat};
	
	\end{semilogyaxis}
	\end{tikzpicture}
	\vspace{-6mm}
	\caption{PAPR reduction performance in SVD-MIMO $ 4 \times 4$ using PTS}
	\label{f7}
\end{minipage}
\vspace{-4mm}
\end{figure*}
\begin{figure*}
\begin{minipage}{0.31\textwidth}
	\begin{tikzpicture}[scale=0.7]
	\begin{semilogyaxis}[
	height=6.5cm,
	width=8.4cm,
	xlabel={$\mathrm{PAPR}_0$ (dB)},
	ylabel={$P(\mathrm{PAPR} > \mathrm{PAPR}_0)$},
	xmin=4, xmax=13, grid=both,
	major grid style={black!40},
	ymin=0.0001, ymax=1.5,
	xtick={4, 5, 6, 7, 8, 9, 10, 11, 12, 13},
	legend entries={No optimization, {$M = 2, U = 2, D = 2$}, {$M = 2, U = 2, D = 4$}, {$M = 2, U = 4, D = 4$}, {$M = 4, U = 2, D = 2$}, {$M = 4, U = 2, D = 4$}, {$M = 4, U = 4, D = 4$}, {$M = 4, U = 4, D = 8$},},
	legend style={cells={anchor=west}, at={(0.99,0.99)}, font=\scriptsize},
	]
	
	\addplot [dashdotdotted,] table {./data/data_svd_4x4.dat};
	\addplot [mark=triangle*, mark options={scale=1,solid},] table {./data/data_svd_4x4_cspts_2p_2w_2s.dat};
	\addplot [densely dashed, mark=diamond*, mark options={scale=1,solid,fill=white},] table {./data/data_svd_4x4_cspts_2p_2w_4s.dat};
	\addplot [mark=pentagon*, mark options={scale=1,solid},] table {./data/data_svd_4x4_cspts_2p_4w_4s.dat};
	\addplot [densely dashed, mark=*, mark options={scale=1,solid,fill=white},] table {./data/data_svd_4x4_cspts_4p_2w_2s.dat};
	\addplot [mark=square*, mark options={scale=1,solid},] table {./data/data_svd_4x4_cspts_4p_2w_4s.dat};
	\addplot [densely dashed, mark=oplus*, mark options={scale=1,solid, fill=white},] table {./data/data_svd_4x4_cspts_4p_4w_4s.dat};
	\addplot [mark=10-pointed star, mark options={scale=1,solid},] table {./data/data_svd_4x4_cspts_4p_4w_8s.dat};
	
	\end{semilogyaxis}
	\end{tikzpicture}
	\vspace{-6mm}
	\caption{PAPR reduction performance in SVD-MIMO $ 4 \times 4$ using CS-PTS}
	\label{f8}
\end{minipage}
\hspace{2mm}
\begin{minipage}{0.31\textwidth}
	\begin{tikzpicture}[scale=0.7]
	\begin{semilogyaxis}[
	height=6.5cm,
	width=8.4cm,
	xlabel={$\mathrm{PAPR}_0$ (dB)},
	ylabel={$P(\mathrm{PAPR} > \mathrm{PAPR}_0)$},
	xmin=6, xmax=7.35, grid=both,
	major grid style={black!40},
	ymin=0.0001, ymax=1.5,
	xtick={6, 6.2, 6.4, 6.6, 6.8, 7.0},
	legend entries={\texttt{Optimal}, 
					{\texttt{CSA}, $\rho=270$}, 
		            {\texttt{SWAN}, $\rho=270$}, 
		            {\texttt{CSA}, $\rho=540$},
		            {\texttt{SWAN}, $\rho=540$},
		            {\texttt{CSA}, $\rho=1080$},
		            {\texttt{SWAN}, $\rho=1080$},
		            {\texttt{CSA}, $\rho=2160$},
		            {\texttt{SWAN}, $\rho=2160$},
		            {\texttt{CSA}, $\rho=4320$},
		            {\texttt{SWAN}, $\rho=4320$}},
	legend style={cells={anchor=west}, at={(0.99,0.99)}, font=\scriptsize},
	]
	
	\addplot [dashdotdotted,]  table {./data/data_svd_4x4_cspts_4p_4w_8s.dat};
	
	\addplot [densely dashed, mark=*, mark options={scale=1,solid, fill=white},] table  {./data/data_csa_270.dat};
	\addplot [mark=oplus*, mark options={scale=1,solid},] table {./data/data_btmcsa_270.dat};
	
	\addplot [densely dashed, mark=triangle*, mark options={scale=1,solid, fill=white},] table  {./data/data_csa_540.dat};
	\addplot [mark=triangle*, mark options={scale=1,solid},] table {./data/data_btmcsa_540.dat};
	
	\addplot [densely dashed, mark=pentagon*, mark options={scale=1,solid, fill=white},] table  {./data/data_csa_1080.dat};
	\addplot [mark=pentagon*, mark options={scale=1,solid,},] table {./data/data_btmcsa_1080.dat};
	
	\addplot [densely dashed, mark=square*, mark options={scale=1,solid, fill=white},] table  {./data/data_csa_2160.dat};
	\addplot [mark=square*, mark options={scale=1,solid},] table {./data/data_btmcsa_2160.dat};
	
	\addplot [densely dashed, mark=diamond*, mark options={scale=1.2,solid, fill=white},] table  {./data/data_csa_4320.dat};
	\addplot [mark=diamond*, mark options={scale=1.2,solid},] table {./data/data_btmcsa_4320.dat};
	
	\end{semilogyaxis}
	\end{tikzpicture}
	\vspace{-6mm}
	\caption{Comparison between \texttt{SWAN} and \texttt{CSA} in SVD-MIMO $ 4 \times 4$}
	\label{f9}
\end{minipage}
\hspace{2mm}
\begin{minipage}{0.31\textwidth}
	\begin{tikzpicture}[scale=0.7]
	\begin{semilogyaxis}[
	height = 6.5cm,
	width = 8.4cm,
	xlabel = {$ \mathrm{PAPR}_0 $ (dB)},
	ylabel = {$ P(\mathrm{PAPR} > \mathrm{PAPR}_0) $},
	xmin = 6, 
	xmax = 7.2, 
	grid = both,
	major grid style = {black!40},
	ymin = 0.0001, 
	ymax = 1.5,
	xtick={6, 6.2, 6.4, 6.6, 6.8, 7.0, 7.2},
	legend entries={\texttt{Optimal}, 
					\texttt{SWAN}, 
					\texttt{CSA},
					\texttt{PSO},
					\texttt{GA}, },
	legend style={cells={anchor=west}, at={(0.99,0.99)}, font=\scriptsize},
	]
	
	\addplot [dashdotdotted,]  table {./data/data_svd_4x4_cspts_4p_4w_8s.dat};

	\addplot [mark=oplus*, mark options={scale=1,solid},] table {./data/data_btmcsa_270.dat};
	\addplot [densely dashed, mark=oplus*, mark options={scale=1,solid, fill=white},] table  {./data/data_csa_270.dat};
	\addplot [mark=diamond*, mark options={scale=1,solid,},] table  {./data/data_pso_270.dat};
	\addplot [densely dashed, mark=triangle*, mark options={scale=1.2,solid, fill=white},] table  {./data/data_ga_270.dat};
	\draw (6.5, 0.001) ellipse(0.4cm and 0.2cm);
	\draw (6.5, 0.00142) -- (6.95, 0.015);
	\node[text width=1.2cm, anchor=west, right] at (6.95, 0.015) {\scriptsize $\rho = 4320$};
	
	\addplot [mark=oplus*, mark options={scale=1,solid},] table {./data/data_btmcsa_1080.dat};
	\addplot [densely dashed, mark=oplus*, mark options={scale=1,solid, fill=white},] table  {./data/data_csa_1080.dat};
	\addplot [mark=diamond*, mark options={scale=1,solid,},] table  {./data/data_pso_1080.dat};
	\addplot [densely dashed, mark=triangle*, mark options={scale=1.2,solid, fill=white},] table  {./data/data_ga_1080.dat};
	\draw (6.73, 0.0005) ellipse(0.4cm and 0.2cm);
	\draw (6.73, 0.00071) -- (6.98, 0.003);
	\node[text width=1.2cm, anchor=west, right] at (6.98, 0.003) {\scriptsize $\rho = 1080$};
	
	\addplot [mark=oplus*, mark options={scale=1,solid},] table {./data/data_btmcsa_4320.dat};
	\addplot [densely dashed, mark=oplus*, mark options={scale=1,solid, fill=white},] table  {./data/data_csa_4320.dat};
	\addplot [mark=diamond*, mark options={scale=1,solid,},] table  {./data/data_pso_4320.dat};
	\addplot [densely dashed, mark=triangle*, mark options={scale=1.2,solid, fill=white},] table  {./data/data_ga_4320.dat};
	\draw (6.99, 0.00025) ellipse(0.4cm and 0.2cm);
	\node[text width=1.2cm, anchor=west, right] at (7.01, 0.0004) {\scriptsize $\rho = 270$};
	
	\end{semilogyaxis}
	\end{tikzpicture}
	\vspace{-6mm}
	\caption{Comparison of swarm-based approaches in SVD-MIMO $ 4 \times 4$}
	\label{f10}
	\end{minipage}
	\vspace{-1mm}
\end{figure*}

\emph{Adaptation of \texttt{SWAN} to CS-PTS in SVD-MIMO}: 
In general, swarm-based approaches cannot be applied straightforwardly. Adjustments are necessary to take into consideration the underlying nature of the problem. In CS-PTS for SVD-MIMO, the number of dimensions is $ n = 2M $, i.e., $ M $ phase rotations and $ M $ time shifts. Without loss of optimality, one phase rotation and one time shift can be fixed since the PAPR changes based on relative phase differences. As a result, the unknown parameters are $ \{ \gamma^\mathrm{opt}_m \}_{m=1}^{M-1} $ and $ \{ \delta^\mathrm{opt}_m \}_{m=1}^{M-1} $. Any candidate solution at iteration $ t $ has the structure $ \mathbf{z}^{(t)}_i = [\gamma^{(t)}_0, \cdots, \gamma^{(t)}_{M-1}, \delta^{(t)}_0, \cdots, \delta^{(t)}_{M-1} ]^T \in \mathbb{X}^n $, with $ \mathbb{X}^n = \underbrace{ \mathbb{C} \times \cdots \times \mathbb{C}}_{n/2} \times \underbrace{ \mathbb{R} \times \cdots \times \mathbb{R} }_{n/2} $, $ {\gamma}^{(t)}_0 = 1 $  and $ {\delta}^{(t)}_0 = 0 $. We define the function $ f: \mathbb{X}^n \to \mathbb{R} $ that takes an $ n $-dimensional input and maps it to a real value, which is the maximum PAPR across all the $ N_\mathrm{tx} $ transmit antennas, i.e., $ f(\mathbf{z}) = \max_{1 \leq i \leq N_\mathrm{tx}} \mathrm{PAPR}_i (\mathbf{z}) $.
As observed in (\ref{e1}), the evaluation of each candidate solution requires $ M-1 $ complex operations due to weighting by $ \gamma_m $ whereas time-shifting by $ \delta_m $ can be accomplished by varying the summation index only. Therefore, most of the computational complexity is due to complex multiplications by $ \gamma_m $. We avoid part of these operations by dividing \texttt{SWAN} into two stages (see Fig. \ref{f4}). To decrease the number of complex multiplications, we create $ \rho_1 $ intermediate nodes which only bear the effect of the phase rotations $ \gamma_m $. From each intermediate node, $ \rho_2 $ solutions bearing the added effect of time-shifting are generated, amounting a total of $ \rho=\rho_1 \rho_2 $ candidates. By adopting the described structure, the complex multiplications we incur into are associated only to $ \rho_1 $ candidate solutions. Finally, the fitness function $ F_\mathbf{z} $ of a candidate solution $ \mathbf{z} $ is defined as $ F_\mathbf{z} = \frac{1}{ 1 + f(\mathbf{z}) } $.

\section{Computational Complexity}
\begin{table*}[!t]
\centering
\caption{Computational complexity of SLM, PTS, CS-PTS using exhaustive search}
\label{t1}
\setlength\tabcolsep{1.5pt}
\begin{tabular}{m{17mm} c c c c c c}
	\toprule
	\centering{Algorithm} & \multicolumn{2}{c}{\centering{SLM}} & \multicolumn{2}{c}{\centering{PTS}} & \multicolumn{2}{c}{\centering{CS-PTS}}\\
	\toprule
	\multirow{2}{17mm}{\centering{Process}} & \multicolumn{2}{c}{Complexity} & \multicolumn{2}{c}{Complexity}	& \multicolumn{2}{c}{Complexity}\\ 
	\cmidrule(r){2-7}
	& Multiplications 	& Additions	& Multiplications 	& Additions	& Multiplications 	& Additions\\ 
	\midrule
	\midrule
	\centering{Zero-padded IFFTs}  & \tiny{$V\left[ \frac{N_cL}{2} \log_2 N_c + \frac{N_cL}{2} \right]$} & \tiny{$V\left[N_cL \log_2 N_c \right]$}
	& \tiny{$M\left[ \frac{N_cL}{2} \log_2 \left(\frac{N_c}{M}\right) + \frac{N_cL}{2} \right]$} & \tiny{$M\left[N_cL \log_2 \left(\frac{N_c}{M}\right) \right]$}	
	& \tiny{$M\left[ \frac{N_cL}{2} \log_2 \left(\frac{N_c}{M}\right) + \frac{N_cL}{2} \right]$} & \tiny{$M\left[N_cL \log_2 \left(\frac{N_c}{M}\right) \right]$}
	\\
	\hdashline
	\centering{Type} & Complex & Complex	& Complex & Complex		& Complex & Complex\\
	\midrule
	\centering{Phase patterns}  & \tiny{$V\left[N_cL\right]$} & \tiny{$0$}	
	& \tiny{$U^{M-1}\left[(M-1)N_cL\right]$} & \tiny{$U^{M-1}\left[(M-1)N_cL\right]$} 
	& \tiny{$U^{M-1}\left[(M-1)N_cL\right]$} & \tiny{$(DU)^{M-1}\left[(M-1)N_cL\right]$}
	\\
	\hdashline
	\centering{Type} & Complex & Complex	& Complex & Complex		& Complex & Complex\\
	\midrule
	\centering{PAPR computation}  & \tiny{$V\left[2N_cL\right]$} & \tiny{$V\left[N_cL\right]$} 
	& \tiny{$U^{M-1}\left[2N_cL\right]$} & \tiny{$U^{M-1}\left[N_cL\right]$}
	& \tiny{$(DU)^{M-1}\left[2N_cL\right]$} & \tiny{$(DU)^{M-1}\left[N_cL\right]$}
	\\
	\hdashline
	\centering{Type} & Real & Real	& Real & Real	& Real & Real\\
	\bottomrule
\end{tabular}
\end{table*}
%
%
%
%
%
%
%
\begin{table}[!t]
	\centering
	\caption{Computational complexity of CS-PTS using \texttt{SWAN} for $\rho=\rho_1 \rho_2$ solution patterns}
	\label{t2}
	\begin{tabular}{m{1.8cm} @{}c @{}c }
		\toprule
		\multirow{2}{1.8cm}{\centering{Process}} & \multicolumn{2}{c}{Complexity} \\ 
		\cmidrule(r){2-3}
		&Multiplications & Additions \\ 
		\midrule
		\midrule
		\centering{Zero-padded IFFTs}  & $M\Big[ \frac{N_cL}{2} \log_2 (\frac{N_c}{M}) + \frac{N_cL}{2} \Big]$ & $M\Big[N_cL \log_2 (\frac{N_c}{M}) \Big]$\\
		\hdashline
		\centering{Type} & Complex & Complex\\
		
		\midrule
		
		\centering{Phase patterns}  & $\rho_1[(M-1)N_cL]$ & $\rho_1 \rho_2[(M-1)N_cL]$\\
		\hdashline
		\centering{Type} & Complex & Complex\\
		
		\midrule
		
		\centering{Generation of solutions}  & $\rho [3(M-1)]$ & $\rho [2(M-1)]$\\
		\hdashline
		\centering{Type} & Real & Real\\
		
		\midrule
		
		\centering{PAPR computation}  & $\rho [2N_cL]$ & $\rho [N_cL]$\\
		\hdashline
		\centering{Type} & Real & Real\\

		\bottomrule
	\end{tabular}
\end{table}
%
%
%
%
%
%
%
%
%
%
%
%
\begin{table}[!t]
	\centering
	\caption{Comparison of computational complexity of swarm-based approaches for $ \rho $ solution patterns}
	\label{t3}
	\begin{tabular}{c m{1.8cm} @{}c @{}c }
		\toprule
		\multicolumn{2}{c} {\multirow{2}{1.8cm}{\centering{Process}}} & \multicolumn{2}{c}{Complexity} \\ 
		\cmidrule(r){3-4}
		&										 & Multiplications & Additions \\ 
		\midrule
		\midrule
		
		\multicolumn{2}{c}{\centering{Zero-padded IFFTs}} & $M\Big[ \frac{N_cL}{2} \log_2 (\frac{N_c}{M}) + \frac{N_cL}{2} \Big]$ & $M\Big[N_cL \log_2 (\frac{N_c}{M}) \Big]$ \\ 
		\hdashline
		\multicolumn{2}{c}{\centering{Type}} & Complex & Complex\\
		\midrule
		
		\multicolumn{2}{c}{\centering{Phase patterns}} & $\rho[(M-1)N_cL]$ & $\rho[(M-1)N_cL]$ \\ 
		\hdashline
		\multicolumn{2}{c}{\centering{Type}} & Complex & Complex\\
		\midrule	
		
		\multicolumn{2}{c}{\centering{PAPR computation}} & $\rho[(M-1)N_cL]$ & $\rho[(M-1)N_cL]$ \\ 
		\hdashline
		\multicolumn{2}{c}{\centering{Type}} & Complex & Complex\\
		\midrule				
		
		\multirow{2}{*}{\rotatebox{90}{\kern-0.3em GA}}  & \centering{Generation of solutions} & $\rho[3(M-1)]$ & $\rho[3(M-1)]$\\
		\cdashline{2-4}
		& \centering{Type} & Complex & Complex\\
		\midrule
		
		\multirow{2}{*}{\rotatebox{90}{\kern-0.3em PSO}}  & \centering{Generation of solutions} & $\rho[5(M-1)]$ & $\rho[5(M-1)]$\\
		\cdashline{2-4}
		& \centering{Type} & Complex & Complex\\
		\midrule
		
		\multirow{2}{*}{\rotatebox{90}{\kern-0.3em CSA}}  & \centering{Generation of solutions} & $\rho[8(M-1)]$ & $\rho[4(M-1)]$\\
		\cdashline{2-4}
		& \centering{Type} & Complex & Complex\\
		
		\bottomrule
	\end{tabular}
\end{table}

Table \ref{t1} shows the complexity of SLM, PTS, and CS-PTS when exhaustive search is employed. The codebook size used by SLM is $ V $, whereas the number of partitions used by either PTS or CS-PTS is $ M $. Also, $ U $ and $ D $ represent the number of admissible phase rotations and time shifts, respectively. As observed, CS-PTS has the highest number of solution patterns (due to the increased dimensionality, i.e., phase rotations and time shifts), which justifies the importance of \texttt{SWAN}. As shown in Table \ref{t2}, when employing \texttt{SWAN} in CS-PTS, the exponential complexity is eliminated and instead it is controlled by $ \rho_1 $ and $ \rho_2 $. Since we compare the performance of \texttt{SWAN} against the benchmarks \texttt{CSA}, \texttt{PSO} and \texttt{GA} in \sref{sec:result}, we also show their complexity in Table \ref{t3}, where $ \rho $ represents the number of generated candidate solutions. Upon comparing Table \ref{t2} and Table \ref{t3}, we conclude that the average cost per generated solution of \texttt{SWAN} is approximately half of that required by \texttt{CSA}. This is a consequence of adding the four mechanisms described in Section \ref{s3}, which require low-complexity operations and, on average, reduce the cost. Complexity is a critical factor when selecting an approach for practicality reasons. Nevertheless, convergence also plays an important role in guaranteeing high performance. For instance, although \texttt{GA} may be a preferred choice over \texttt{CSA} and \texttt{PSO} due to its low cost per generation, we corroborate in the next section that \texttt{GA} performs worst in terms of convergence (i.e., for a given number of generations the performance of \texttt{GA} is subpar compared to \texttt{CSA} and \texttt{PSO}, thus exhibiting its lower convergence per iteration).

\section{Numerical Experiments}
\label{sec:result}

In this section, we evaluate the probability that the PAPR exceeds a threshold $ \mathrm{PAPR}_0 $, denoted by $ P(\mathrm{PAPR}>\mathrm{PAPR}_0) $. We evaluate several techniques under various configurations. For a fair comparison, we also adapt SLM and PTS to operate in SVD-MIMO mode. In the sequel, we assume that the data symbols are randomly obtained from a 64-QAM constellation, the oversampling factor is $ L = 4 $, and the number of subcarriers is $ N_c = 256 $. We also assume the Rayleigh fading channel model \cite{abanto2020:hydrawave-multigroup-multicast-hybrid-precoding-low-latency-scheduling, stuber2004:broadband-mimo-ofdm-wireless-communications} with $ N_p = 16 $ paths.

Fig. \ref{f5} shows the PAPR for a varying number of transmit and receive antennas ($ N_\mathrm{tx} \times N_\mathrm{rx} $) when PAPR reduction is not considered. As the number of antennas increases, the \emph{min-max} PAPR (evaluated via (\ref{e1})) increases as well. This is an expect outcome since a given solution needs to minimize the maximum PAPR over multiple antennas. In the following scenarios (i.e., Fig. \ref{f6} to Fig. \ref{f10}), we evaluate a variety of PAPR reduction techniques when considering a $ 4 \times 4 $ MIMO system with SVD precoding. 

Fig. \ref{f6} shows $ P(\mathrm{PAPR}>\mathrm{PAPR}_0) $ using SLM with different $ V = \{4, 16, 64, 256, 1024, 4096\} $, which are pseudo-randomly generated with phase rotations from $ \left\lbrace e^0, e^{\pi} \right\rbrace $. Fig. \ref{f7} shows the performance of PTS with $ M =\left\lbrace 2, 4, 8 \right\rbrace $, $ U = \left\lbrace 2, 4, 8 \right\rbrace $ whereas Fig. \ref{f8} shows the performance of CS-PTS with $ M =\left\lbrace 2, 4, 8 \right\rbrace $, $ U = \left\lbrace 2, 4, 8 \right\rbrace $, $ D = \left\lbrace 2, 4, 8 \right\rbrace $. Considering the trade-off between complexity and performance, CS-PTS achieves superior results compared to PTS and SLM. Specifically, CS-PTS generates several solution patterns by solely time-shifting the partial transmit sequences, which does not incur in additional costly complex multiplications. Fig. \ref{f9} depicts the performance attained by \texttt{CSA} and \texttt{SWAN}, as well as $\texttt{Optimal}$ (i.e., obtained through exhaustive search) when $ M = 4 $, $ U = 4 $, $ D = 8 $. For \texttt{CSA} and \texttt{SWAN}, we consider a variety of iterations $ \rho = \left\lbrace 270, 540, 1080, 2160, 4320 \right\rbrace $. Although \texttt{CSA} is computationally more complex than \texttt{PSO} and \texttt{GA} (as seen in Table \ref{t3}), we consider \texttt{CSA} as the benchmark approach due to its higher performance in terms of optimality. With almost half of the complexity of \texttt{CSA} (compare Table \ref{t2} and Table \ref{t3}), \texttt{SWAN} consistently outperforms \texttt{CSA} under the same number of iterations. Also, the results shown under $ \texttt{Optimal} $ are obtained after evaluating $ (4\times8)^3 = 32,768 $ solution candidates. We realize that by only evaluating $ \rho = 4320 $ patterns, \texttt{SWAN} is at most $ 0.1 $ dB apart from $ \texttt{Optimal} $ with a probability of $10^{-4}$. Fig. \ref{f10} shows that \texttt{CSA} outperforms \texttt{PSO} and \texttt{GA}, and has higher convergence rate. Noteworthily, under the same $ \rho $ value, \texttt{SWAN} always outperforms \texttt{CSA}, \texttt{PSO}, and \texttt{GA}. Although \texttt{SWAN} and \texttt{GA} have comparable complexities, \texttt{SWAN} outperforms \texttt{GA} by $ 0.15 $ dB at $ P(\mathrm{PAPR} > \mathrm{PAPR}_0) = 10^{-4} $, and this result is prevailing under all the evaluated values of $ \rho = \left\lbrace 270, 1080, 4320 \right\rbrace $. 
     

\section{Conclusions}
In this paper, we adapted CS-PTS to operate in MIMO systems with SVD precoding. Leveraging on this system, we formulated a \emph{min-max} problem to reduce the PAPR across multiple transmit antennas. Given the high computational complexity of the resulting problem, we proposed a swarm-based approach called \texttt{SWAN} to design the parameters (i.e., phase rotations and time shifts) that minimize the maximum PAPR. Through extensive simulations, we showed that \texttt{SWAN} outperforms other competing approaches such as \texttt{CSA}, \texttt{GA}, and \texttt{PSO} in terms of convergence and complexity. Our results confirmed that even with a low complexity requirement, \texttt{SWAN} attains near-optimality. We conclude that \texttt{SWAN} is an attractive technique for systems with limited capabilities. In particular, through \texttt{SWAN}, computationally-constrained systems can explore the solution space in a smarter fashion, thus providing a better trade-off between complexity and optimality compared to straightforward approaches such as exhaustive search.

\section*{Acknowledgment}
This research is funded by the Deutsche Forschungsgemeinschaft (DFG) within the B5G-Cell project in SFB 1053 MAKI, and the LOEWE initiative (Hesse, Germany) within the emergenCITY centre.

\bibliographystyle{IEEEtran}
\bibliography{ref}

\end{document}